\documentclass{cjaa}                   
\usepackage{graphicx}                   
\input{epsf.sty}                        
\input{psfig.sty}                       

\begin{document}

   \title{Monopole-charged pulsars and relevant issues}

   \volnopage{Vol.0 (200x) No.0, 000--000}      
   \setcounter{page}{1}          

   \author{Xiao-Hong Cui, You-Ling Yue, Ren-Xin Xu, Guo-Jun Qiao
      }
   \offprints{Xiao-Hong Cui}                   

   \institute{Astronomy Department, School of Physics, Peking University, Beijing 100871, China\\
             \email{xhcui@bac.pku.edu.cn}
          }

   \date{Received~~~~~~~~~~~~~~~~~; ~~accepted}

   \abstract{
The {\em aligned} pulsars whose rotation axes and magnetic dipole
axes are parallel should be positively charged. The total charge
of pulsars is calculated after considering the electromagnetic
field in and out the star under a specific condition.
The statistical relation between the pulsar's rotation energy loss
rate (or the period derivative) and the period may hint that the
millisecond radio pulsars with small periods could be low-mass
bare strange stars.
   \keywords{pulsars: general --- stars: neutron --- dense matter}
   }

   \authorrunning{Cui, Yue, Xu \& Qiao}            
   \titlerunning{The Current Flows in Pulsar Magnetospheres}  

   \maketitle
%
\section{Introduction}

The study of pulsar magnetosphere is essential for us to
understand various radiative processes, and thus observed emission
in different bands. The charge-separated plasma (Goldreich and
Julian 1969) in the magnetosphere of various gap models (Ruderman
\& Sutherland 1975; Arons \& Scharlemenn 1979; Cheng, Ho, \&
Ruderman 1986; Qiao et al. 2004) has been investigated.

After investigating the electromagnetic field of an {\em aligned}
pulsar whose rotation axis is parallel to the magnetic axis, Xu,
Cui \& Qiao (2006) suggested that these pulsars should be
positively monopole-charged. We would like to study further
charged pulsars in this paper. Supposing the pulsar is magnetized
homogenously, we calculate the total charges $Q$. Provided that
this charge is proportional to the radiation of pulsars, $Q$
should be proportional to $\dot{E}$, the spin down energy loss
rate. The statistical relation between the pulsar's $\dot{E}$ (or
the period derivative $\dot{P}$) and the period may hint that the
millisecond radio pulsars with small periods could be low-mass
bare strange stars.

This paper is arranged as follows. The total charge of pulsars is
estimated in \S2. In \S3, we show possible evidence for low-mass
millisecond pulsars with small $P$ by observational data.
Conclusions are presented in \S4.

\section{Pulsar's charge estimated}

If the pulsar is assumed to be magnetized homogenously, its
magnetic field could be described as
\begin{equation}
\vec{B}_{\rm in}=\frac{8\pi}{3}\vec{M}=\frac{2\vec{m}}{r_0^3}~~~~(r<r_0),
\label{Bi}
\end{equation}
and
\begin{equation}
\vec{B}_{\rm out}=\frac{3\hat{r}(\hat{r}\cdot
\vec{m})-\vec{m}}{r^3}~~~(r>r_0), \label{Bo}
\end{equation}
where $\vec{M}$ is the magnetized intensity, $\vec{m}$ is the
magnetic dipole moment, $r_0$ is the radius of pulsar, $r$ is the
distance from one arbitrary point in space to the stellar core,
$\hat{r}$ is the corresponding unit vector. Here we define
$|\vec{B}_{\rm in}|\equiv B_0$.

If all the particles in and out the pulsar rotate synchronously
around the magnetic axis, the electric field, $\vec{E}$, could
satisfy
\begin{equation}
\vec{E}+\frac{\vec{\Omega}\times \vec{r}}{c} \times \vec{B}=0,
\label{E}
\end{equation}
without considering the magnetic field induced by synchronously
rotating particles, where $\vec{\Omega}$ is the angular velocity
of star rotating around the dipole rotation axis, which relates
with rotating period of star {\em P} by $P=2\pi/\Omega$. If
$\hat{k}$ is the unit vector of angular momentum,
$\vec{\Omega}=\Omega \hat{k}$ and $\vec{m}=m\hat{k}$ for {\em
aligned} pulsars. We estimate the charge under this specific
condition.

From Eqs.(\ref{Bi}) to (\ref{E}), the electric fields in and out a
pulsar are, respectively,
\begin{equation}
\vec{E}_{\rm in}=\frac{\Omega r
B_0}{c}\hat{k}\times(\hat{k}\times\hat{r})~~~(r<r_0), \label{Ei}
\end{equation}
\begin{equation}
\vec{E}_{\rm out}=\frac{\Omega r_0^3 B_0}{2cr^2}[3(\hat{k}\cdot
\hat{r})\hat{r}-\hat{k}]\times(\hat{k}\times\hat{r})~~~(r>r_0).
\label{Eo}
\end{equation}
The corresponding charges then could be
\begin{equation}
Q_{\rm in} =\rho_{\rm in}\frac{4\pi r_0^3}{3} =\frac{\nabla\cdot
\vec{E}_{\rm in}}{4\pi}\frac{4\pi r^{3}_{0}}{3}
=-\frac{2r_0^3\Omega B_{0}}{3c},
\end{equation}
and $Q_{\rm out} =\int \rho_{\rm out}\cdot dV=0$, where $\rho_{\rm
in}$ and $\rho_{\rm out}$ are the charge densities in and out the
star.

From Eqs.(\ref{Ei}) and (\ref{Eo}), the charge intensity and the
charges on the stellar surface is
\begin{equation}
\sigma_{\rm s}=\frac{1}{4\pi}(\vec{E}_{\rm out}-\vec{E}_{\rm
in})_{r=r_0}\cdot\hat{r}=\frac{3 \Omega r_0 B_0 \sin^2\theta}{8\pi
c},
\end{equation}
\begin{equation}
Q_{\rm s} =\int_{r=r_0}\sigma_{\rm s}\cdot ds=\frac{r_0^3\Omega
B_{0}}{c},
\end{equation}
where $\theta$ is the polar angular between arbitrary vector
$\vec{r}$ and $\hat{k}$.

From the calculation above, we can obtain the total charges
\begin{equation}
Q_{\rm total}=Q_{\rm in}+Q_{\rm out}+Q_{\rm s}=\frac{2\pi R_0^3
B_0}{3Pc}\simeq 2.3\times10^{10}\frac{r_{6}^3 B_{12}}{P}~~({\rm
coulombs}), \label{QQ}
\end{equation}
where $r_{6}=r_0/(10^6$ cm), $B_{12}=B_0/(10^{12}$ G). From this
equation, it is shown that the aligned pulsars should be
positively charged.

It is evident that this total charge in Eq. (\ref{QQ}) is
different from that obtained in Eq. (17) of Xu et al. (2006,
$Q\sim 10^{-3}r_{6}^3 B_{12}/P^2$).
This difference could result in a shift of the boundary (i.e., the
field lines being at the same electric potential as the
interstellar medium) which separating regions I and II in Fig. 1
of Xu et al. (2006): the boundary should be closer (i.e., above
the critical field lines) to the magnetic axis if the real charge
is greater than $Q$.
Unfortunately, we are faraway from obtaining an exact value of
charge since a global solution to pulsar magnetosphere is still
impossible.

\section{Evidence for low-mass millisecond pulsars}

There are 1394 observational
pulsars\footnote{http://www.atnf.csiro.au/research/pulsar/psrcat/}
with known {\em P} and $\dot{P}$ (thus $\dot{E}$) simultaneously.
The numbers of millisecond and normal radio pulsars are 87 and
1307 if the dividing lines is by an ``apparent'' magnetic field of
$B_0=6.4\times10^{19}\sqrt{P\dot{P}}$ G$=5\times10^{10}$ G as
shown in left upper panel of Fig.1. Here we only consider 87
millisecond pulsars ($B<B_0$ G).

\begin{figure}
  \centering
    \includegraphics[width=14cm]{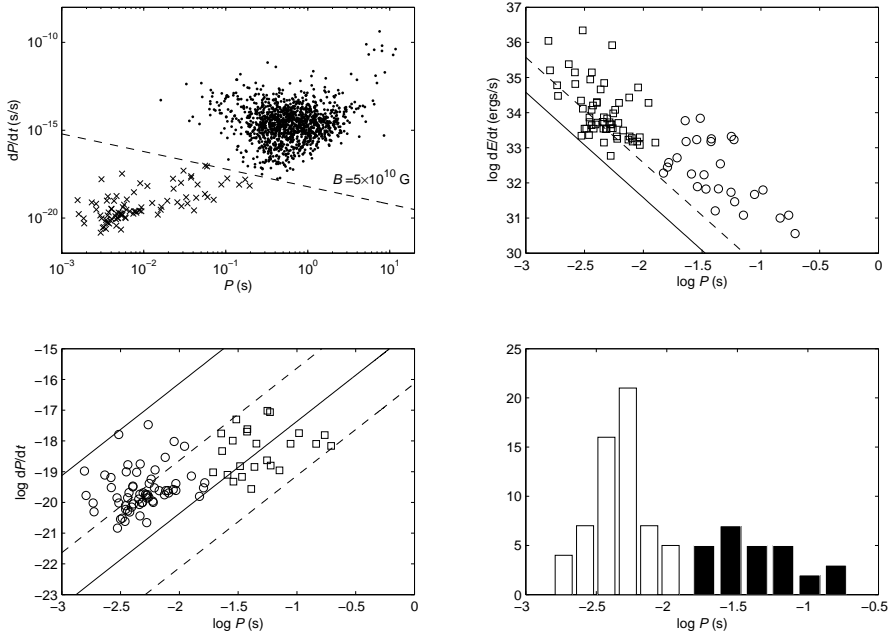}
    \caption{Left upper panel:
Period derivatives $\dot{P}$ as a function of period $P$ for 1349
radio pulsars. The dots represent the normal radio pulsars with
$B>B_0$ and the stars are the millisecond radio pulsars with
$B<B_0$ if the dividing line of them is the magnetic field
$B_0=5\times10^{10}$ G. Left lower panel: The period derivative
$\dot{P}$ as a function of period $P$ for millisecond radio
pulsars. The empty circles ({\em Group I'}) and squares ({\em
Group II'}) are two groups divided by Kmeans methods. The regions
between two solid lines(for {\em Group I'}) and dash lines ({\em
Group II'}) represent limit ranges of radii for these two groups,
respectively. Right upper panel: Spin down energy loss rate
$\dot{E}$ as a function of period $P$ for millisecond radio
pulsars. The empty circles ({\em Group I}) and squares ({\em Group
II}) are two groups divided by Kmeans methods. The solid and dash
lines represent the minimums of radius for these two groups. Right
lower panel: Two-peak period $P$ distribution of 87 millisecond
radio pulsars.}
\end{figure}

It is known that the numerical value of rotation energy loss rate
\begin{equation}
\dot E=-I \Omega{\dot \Omega}\simeq 3.1\times10^{25}(\frac{r_{6}}{10^6})^2\frac{\dot{P}_{-20}}{10^{-20}}P^{-3},%
\label{Edot}
\end{equation}
where rotational inertia of star $I\simeq \frac{2}{5}Mr_{6}^2$
(the star is assumed to be a homogeneous rigid sphere),
$M=M_\odot$, $\dot{P}_{-20}=\dot{P}/10^{-20}$. Note that the
pulsar radius could be a variable. A potential drop across the
open field line region for a rotating magnetic pulsar could be
simply
\begin{equation}
\Phi=6.6\times10^{12}B_{12}r_{6}^3P^{-2}~~~~\rm V.%
\label{Phi}
\end{equation}

For giving the hints about the star radius and showing possible
evidence for low-mass millisecond radio pulsars in the bare
strange star model by observational data, we investigate the
relations of $\dot{E}-P$ and $\dot{P}-P$ as shown in left lower
panel and right upper panel of Fig.1. Secondly, applying the
Kmeans method, we divide them into two groups: 60 pulsars in {\em
Group I} and 27 pulsars in {\em Group II}; 63 in {\em Group I'}
and 24 in {\em Group II'}. Thirdly, from
$B=6.4\times10^{19}\sqrt{P\dot{P}}$ G and assign
$\dot{P}_{-20}=1.0$, $\Phi=1.0\times10^{12} V$ in Eqs.(\ref{Edot})
and (\ref{Phi}) for 87 millisecond pulsars, we give the minimums
of radius for these two groups $r_{6I}\approx0.35$ and
$r_{6II}\approx1.1$ and the limit ranges of radii $0.065\leq
r_{6I'}\leq 0.35$ and $0.17\leq r_{6II'}\leq 0.65$. Finally, from
the analysis above and according to Eq.(10) in Xu (2005), we can
obtain the radius and mass ratios of {\em
Group I} to {\em Group II} 
from the relation of $P-\dot{E}$ and those 
of {\em Group I'} to {\em Group II'} from
the relation of $P-\dot{P}$ as shown in Table 1.

\begin{table}[]
  \caption[]{The radius and mass ratios of {\em Group I} to {\em Group II} from the relation of $P-\dot{E}$ and
  that of {\em Group I'} to {\em Group II'} from the relation of $P-\dot{P}$.} 
%
  \label{Tab:publ-works}
  \begin{center}\begin{tabular}{cccccc}
  \hline\noalign{\smallskip}
Relation &  Radius Radio  & Mass Ratio     \\
  \hline\noalign{\smallskip}
$P-\dot{E}$ & 0.32  & 0.03    \\
$P-\dot{P}$ &  0.01 & 0.13  \\
  \noalign{\smallskip}\hline
  \end{tabular}\end{center}
\end{table}

\section{Conclusions}

The total charge of an uniformly magnetized aligned pulsar with
Goldreich-Julian density outside the star is calculated.
The boundary which separates the positive and negative flows in a
magnetosphere is not necessary the critical lines if the real
charge is not that given by Xu et al. (2006).
The statistical relation between the pulsar's $\dot{E}$ (or the
period derivative $\dot{P}$) and the period $P$ may indicate that
the millisecond radio pulsars with small periods could be low-mass
bare strange stars.


{\em Acknowledgments}:
The authors thank helpful discussion with the members in the
pulsar group of Peking University. This work is supported by NSFC
(10273001, 10573002), the Special Funds for Major State Basic
Research Projects of China (G2000077602), and by the Key Grant
Project of Chinese Ministry of Education (305001).


\end{document}